\documentclass[a4paper,10pt,oneside]{article}
\usepackage[english]{babel}
\usepackage{t1enc}
\usepackage[latin2]{inputenc}
\usepackage{graphics}
\usepackage{psfrag}
\usepackage[a4paper,left=2.5cm,right=2.5cm,top=2cm,bottom=2cm]{geometry}
\usepackage{amsmath}
\usepackage{amssymb}
\usepackage{amscd}
\usepackage{array}
\usepackage{enumerate}
\usepackage{tabularx}
\usepackage[mathscr]{eucal}
\usepackage{graphicx}
\usepackage{hhline}
\usepackage{wrapfig}
\usepackage{rotating}
\usepackage{verbatim}
\usepackage{ntheorem}
\usepackage{float}
\usepackage{subfig}
\usepackage{url}
\usepackage{authblk}
\usepackage{xcolor,colortbl}
\usepackage{multirow}

\usepackage{eso-pic}
\usepackage{everyshi}
\usepackage{ifthen}
\usepackage{calc}
\usepackage[square, sectionbib]{natbib}

\usepackage[colorlinks=false, bookmarks=true]{hyperref}

\title{Comparison of Stochastic Claims Reserving Models in Insurance}
\author[1]{\textsc{L\'{a}szl\'{o} Martinek} \thanks{martinek@cs.elte.hu $\bullet$ Room D-3-309, P\'{a}zm\'{a}ny P\'{e}ter s\'{e}t\'{a}ny 1/C, 1117 Budapest, Hungary}}
\author[1]{\textsc{Mikl\'{o}s Arat\'{o}} \thanks{aratonm@ludens.elte.hu $\bullet$ Room D-3-311, P\'{a}zm\'{a}ny P\'{e}ter s\'{e}t\'{a}ny 1/C, 1117 Budapest, Hungary}}
\author[1]{\textsc{Mikl\'{o}s M\'{a}lyusz}}
\affil[1]{Institute of Mathematics, E\"{o}tv\"{o}s Lor\'{a}nd University, Budapest, Hungary}

\begin{document}

\maketitle
\numberwithin{figure}{section}
\numberwithin{equation}{section}
\numberwithin{table}{section}

\pagenumbering{arabic}
\setcounter{page}{1}

\renewcommand{\contentsname}{Table of Contents}
\tableofcontents*
\setcounter{tocdepth}{2}


\renewcommand{\abstractname}{Abstract}
\begin{abstract}
The appropriate estimation of incurred but not reported (IBNR) reserves is traditionally one of the most important task of actuaries working in casualty and property insurance. As certain claims are reported many years after their occurrence, the amount and appropriateness of the reserves has a strong effect on the results of the institution. In recent years, stochastic reserving methods had become increasingly widespread. The topic has a wide actuarial literature, describing development models and evaluation techniques. We only mention the summarizing article \cite{EV2002} and book \cite{MV2008}.

The cardinal aim of our present work is the comparison of appropriateness of several stochastic estimation methods, supposing different distributional development models. We view stochastic reserving as a stochastic forecast, so using the comparison techniques developed for stochastic forecasts, namely scores, is natural, see \cite{GT2007}, for instance. We expect that in some cases this attitude will be more informative than the classical mean square error of prediction measure.

\end{abstract}


\section[Introduction]{Introduction}
\label{sec:intro}

The appropriate estimation of outstanding claims (incurred but not reported (IBNR) claims and reported but not settled (RBNS) claims) is crucial to preserve the solvency of insurance institutions, especially in casualty and property insurance. A general assumption is that claims related to policyholders, which occur in a given year, are reported to the insurance company in the subsequent years, sometimes many years later. Often, the payout is delayed as well. Thus, reserves have to be made to cover these arising obligations.

Data are usually represented by so called run-off triangles. These are $n \times m$ matrices, where element $X_{i,j}$ ($i,j = 1,2,\ldots$) represents the claim amount incurred in year $i$, and paid with a delay of $j-1$ years, or may also indicate aggregate values. In our research work we suppose that $n=m$, i.e., restrict data to be squared in that sense, for simplicity reasons. Although, thereby generality will not be violated, methods can be similarly used for arbitrary $n$ and $m$ positive integers.

Elements for indices $i+j > n+1$ are unknown, have to be predicted. If $X_{i,j}$ denotes an incremental value in the triangle, then the outstanding claims are $\sum \limits_{1 \leq i,j, \leq n} X_{i,j} - \sum \limits_{i+j \leq n+1} X_{i,j}$. We focus on the so called ultimate claim value ($UC$), which is the sum of observed payments (upper triangle) and outstanding claims (lower triangle). Remark that throughout the entire article, $X_{i,j}$ will denote incremental, and $C_{i,j}$ will denote aggregate claim values, i.e., $C_{i,j} = X_{i,1} + \ldots + X_{i,j}$.

The topic has a wide actuarial literature, describing development models and evaluation techniques. See \cite{T2000, M1994, V1994}, for instance. After the quintessential works of these authors, stochastic claims reserving methods are becoming more and more common. These methods not only estimate the expected value of the outstanding claims, but also examine its stochastic behavior. An extensive description of stochastic reserving can be found in the summarizing article \cite{EV2002} and book \cite{MV2008}.

Most of the stochastic reserving methods estimate the distribution of the amount of outstanding claims, so they can be considered as probabilistic forecasts. There are numerous methods for the comparison of probabilistic forecasts, and we chose to concentrate on the probabilistic scores, see \cite{GT2007}.
The cardinal aim of our present work is to demonstrate that using these comparisons is an adequate way of comparing stochastic reserving methods. Our hope is that in some cases this attitude will be more informative than the classical mean square error of prediction measure or quantile values. We also will pay attention to the sharpness of predictions. The goal of our work is the comparison of appropriateness of several stochastic estimation methods, supposing different distributional development models. We used several run-off triangles from the actuarial literature. It is important to note that the results we present are not fully comprehensive, as it would far exceed the constraints of this article. For example, we only used paid run-off triangles, thus we could not make comparisons with the MCMC model described in \cite{MV2010}. Due to lack of capacity, we did not calculate with the bootstrap methods in \cite{BHO2009} and \cite{PSC2003}, and we left out the MCMC method from \cite{GM2007}.

Of the scores, we used only the CRPS and the Energy score, since in most cases a Monte Carlo type evaluation is needed, as the explicit description of the distribution of the ultimate claim value is not possible or too difficult. For instance, this is the case if this value is the sum of random variables derived from log-normal distribution. We applied our comparison method for an itemized (claims and payouts) dataset from an insurance company, where both the upper and the lower triangles were known. We created 2000 scenarios with random draw and replacement from the claims, and chose the most appropriate stochastic reserving methods based on these scenarios.

\section[Distributional Models]{Distributional Models}
\label{sec:model}

In the following section we briefly expound the distributional models used in our research.

\subsection{Log-normal Model}

The following calculation is based on \cite[p. 168.]{MV2008}. Briefly, in the Log-normal model, triangular elements $C_{i,j}$ ($i,j \in \mathbb{Z}_+, ~ i+j \leq n+1$) indicate cumulative claims. The so called $F_{i,j} = \frac{C_{i,j}}{C_{i,j-1}}$ development factors are log-normally distributed random variables, with $\mu_j$ log-scale and $\sigma_j^2$ shape parameters, and let $C_{i,0}$ be 1. Here we only remark that the parameter estimation is given by the solution of Equation \ref{eq:lognormmu}, \ref{eq:lognormsigma}, except $\hat{\sigma_n} := 0$. For further details see \cite{MV2008}.

\begin{equation}\label{eq:lognormmu}
\hat{\mu_j} = \frac{1}{n-j+1} \sum \limits_{i=1}^{n-j+1} \log \left( \frac{C_{i,j}}{C_{i,j-1}} \right) ~~~~~ j \in \{1,\ldots, n \}
\end{equation}
\begin{equation}\label{eq:lognormsigma}
\hat{\sigma^2_j} = \frac{1}{n-j} \sum \limits_{i=1}^{n-j+1} \left( \log \left( \frac{C_{i,j}}{C_{i,j-1}} \right) - \hat{\mu_j} \right)^2 ~~~~~ j \in \{ 1, \ldots, n-1 \}
\end{equation}

Obviously, the distributions of $C_{i,n}$ values -- in other words the ultimate payments for accident year $i$ -- are also log-normal with log-scale parameter $\sum \limits_{k=1}^n \mu_k$, and shape parameter $\sum \limits_{k=1}^n \sigma^2_k$. Unfortunately, the distribution of $UC = \sum \limits_{i=1}^n C_{i,n}$ cannot be expressed explicitely, since this is the sum of $n$ log-normal variables.

\subsection{Negative Binomial Model}

The construction of negative binomial development model below is based on \cite[p. 183.]{MV2008}. Rows are assumed to be independent, and in each row, the distribution of $X_{ij}$ increment value is Poisson$\left( \Theta_{i,j-1} \cdot \left( \frac{\beta_j}{\beta_{j-1}} - 1 \right) \right)$. Here, $\Theta_{i,j-1} \sim \Gamma(c_{i,j-1},1)$ random variable under the assumption that $\{ c_{i,j} = C_{i,j} \}$, thus, the model is recursive. On the other hand, for $j \in \{ 1, \ldots, n \}$, $\beta_j = \sum \limits_{k=1}^j \gamma_k$ are the partial sums of $\underline{\gamma} = (\gamma_1, \ldots, \gamma_n)$ payout pattern. Let $f_j := \frac{\beta_j}{\beta_{j-1}}$ $(j=1,\ldots,n-1)$, and suppose that $f_j > 1$, i.e., $\gamma_k > 0$ for all $k$, which means an increasing payout pattern. It can be shown that the distribution of increment $X_{ij}$ -- conditionally on $C_{i,j-1}$ -- is NegBinom$\left( C_{i,j-1}, \frac{1}{f_{j-1}} \right)$ ($j=2,\ldots,n$). Remark that the NegBinom$(r,p)$ distribution is defined as $P(\xi = k) = { {r+k-1}\choose{k} } p^r (1-p)^k$ $(k=0,1,\ldots)$.

Parameter estimation is based on maximum likelihood estimator as follows. Let $\underline{\gamma}$ denote the unknown parameters, $X$ the upper triangle, moreover, suppose that the first column is fixed. The likelihood function is
$$L(\underline{\gamma},X) = \prod \limits_{i=1}^{n-1} P(row_i) = \prod \limits_{i=1}^{n-1} P(X_{i,2} = k_{i,2} , \ldots, X_{i,n-i+1} = k_{i,n-i+1}) =$$
$$\prod \limits_{i=1}^{n-1} P(X_{i,n-i+1} = k_{i,n-i+1} | X_{i,n-i} = k_{i,n-i}, \ldots, X_{i,2} = k_{i,2}, X_{i,1} = k_{i,1}) \cdot \ldots $$ $$ \ldots \cdot P(X_{i,2} = k_{i,2} | X_{i,1} = k_{i,1}) =$$
$$\prod \limits_{i=1}^{n-1} {C_{i,n-i} + k_{i,n-i+1}-1 \choose k_{i,n-i+1}} p_{n-i}^{C_{i,n-i}} (1-p_{n-i})^{k_{i,n-i+1}} \cdot \ldots$$ $$\ldots \cdot {C_{i,1} + k_{i,2}-1 \choose k_{i,2}} p_{1}^{C_{i,1}} (1-p_{1})^{k_{i,2}} =$$
$$\prod \limits_{j=1}^{n-1} \prod \limits_{i=1}^{n-j} {C_{i,j} + k_{i,j+1} - 1 \choose k_{i,j+1}} p_j^{C_{i,j}} (1-p_j)^{k_{i,j+1}} =$$
$$const \cdot \prod \limits_{j=1}^{n-1} p_j^{\sum \limits_{i=1}^{n-j} C_{i,j}} (1-p_j)^{\sum \limits_{i=1}^{n-j} k_{i,j+1}} \longrightarrow \max \limits_{\underline{p}}.$$

Let $l(\underline{\gamma}, X)$ be $\log L(\underline{\gamma}, X)$, i.e., the loglikelihood:
$$l(\underline{\gamma}, X) = const + \sum \limits_{j=1}^{n-1} \sum \limits_{i=1}^{n-j} C_{i,j} \log p_j + \sum \limits_{j=1}^{n-1} \sum \limits_{i=1}^{n-j} k_{i,j+1} \log(1-p_j).$$

Thus we get the following system of equations: $\frac{\partial}{\partial p_j} l = \frac{\sum \limits_{i=1}^{n-j} C_{i,j}}{p_j} - \frac{\sum \limits_{i=1}^{n-j} k_{i,j+1}}{1-p_j} = 0$, $j \in \{1, \ldots, n-1 \}$. Suppose that $\forall j$ $\sum \limits_{i=1}^{n-j} k_{i,j+1} > 0$. This gives the estimator $$\hat{p}_j = \frac{\sum \limits_{i=1}^{n-j} C_{i,j}}{\sum \limits_{i=1}^{n-j} C_{i,j+1}},$$ which relate to the chain-ladder  development factors.
Note that $\gamma_1 = p_1 p_2 \ldots p_{n-1}$, $\gamma_i = (1-p_{i-1})p_i \ldots p_{n-1}$ $(i = 2,\ldots, n-1)$, $\gamma_n = 1-p_{n-1}$.

At last, note that in \cite{EV2002} another approach of the Negative Binomial Model can be found. According to that paper, a dispersion parameter is included, and increments are so called over-dispersed negative binomially distributed random variables with mean $(f_j - 1)\cdot C_{i,j-1}$ and variance $\phi f_j (f_j - 1) \cdot C_{i,j-1}$, respectively.

\subsection{Poisson Model}

The Poisson model is available in \cite[p. 25.]{MV2008}, for instance. Briefly, suppose that the increments are independent $X_{ij} \sim Poisson(\mu_i \gamma_j)$ variables, $i,j = 1, \ldots, n$. Now suppose that the upper triangle - as a condition, denoted by $\mathcal{D}$ - is given. On the one hand, the estimation of $\gamma_1, \ldots, \gamma_n$ payout pattern values works exactly the same way as in the Negative Binomial case. On the other hand, $$\hat{\mu}_i = \frac{\sum \limits_{k=1}^{n-i+1} X_{ik}}{\sum \limits_{k=1}^{n-i+1} \hat{\gamma}_k}.$$ Since one of our aims is to generate ultimate claim values, assuming that the real parameters are $\hat{\gamma}_i$ and $\hat{\mu}_i$, one random ultimate claim variable is the sum of upper triangle values (or $\sum \limits_{i=1}^{n} C_{i,n-i+1}$) and $Y$ lower triangle, where $Y \sim$ Poisson $\left( \sum \limits_{i=2}^n \hat{\mu}_i (\hat{\gamma}_{n-i+2} + \ldots + \hat{\gamma}_n) \right)$, i.e., $UC = \sum \limits_{i=1}^{n} C_{i,n-i+1} + Y$.

\subsection{Over-dispersed Poisson Model}

A slightly more general model rather applied in practice is the Poisson model with a dispersion parameter. We have the following distributional assumptions. If $Y_{ij} \sim Poisson \left( \frac{\mu_i \gamma_j}{\phi} \right)$ random variable, let $X_{ij} := \phi Y_{ij}$. Thus $E(X_{ij}) = \mu_i \gamma_j$ and $Var(X_{ij}) = \phi \mu_i \gamma_j$, and $X_{ij}$ is called over-dispersed Poisson random variable.

For $\underline{\mu}$ and $\underline{\gamma}$ we used the regular chain-ladder method, which provides unbiased estimation. On the other hand, evaluation of $\phi$ requires the determination of Pearson residuals, for instance. For a detailed description see \cite[p. 218.]{MV2008}, \cite{EV2006}. Briefly, in the over-dispersed Poisson case, Pearson residuals are defined as
$$\hat{r}_{ij}^P = \frac{x_{ij} - \hat{m}_{ij}}{\sqrt{\hat{m}_{ij}}} = \frac{x_{ij} - \hat{\mu}_i \hat{\gamma}_j}{\sqrt{\hat{\mu}_i \hat{\gamma}_j}}.$$
Thus
$$\hat{\phi}_P = \frac{ \sum \limits_{i+j \leq n+1} \left( \hat{r}_{ij}^P \right)^2 }{N-p},$$
where $N$ denotes the number of observations, i.e., $N=\frac{n(n+1)}{2}$, and $p$ is the number of predicted unknown parameters, i.e., $p=2n-1$. Remark that this method provides a biased estimator for $\phi$ (and also for $\underline{\mu}$ parameters).

The distribution of $UC$, similarly to the Poisson model is $UC = \sum \limits_{i=1}^{n} C_{i,n-i+1} + Y$, where $Y \sim \phi \cdot Poisson \left( \frac{1}{\phi} \sum \limits_{i=2}^n \hat{\mu}_i (\hat{\gamma}_{n-i+2} + \ldots + \hat{\gamma}_n) \right)$

Remark that another parameterization of the model is as follows. Let $\alpha_1, \ldots, \alpha_n;$ $\beta_1, \ldots, \beta_n; c; \phi$ be parameters. Incremental claims are defined as \\$X_{ij} \sim \phi \cdot Poisson \left( \frac{e^{\alpha_i + \beta_j + c}}{\phi} \right)$. Thus $E(X_{ij}) = e^{\alpha_i + \beta_j + c}$ and $Var(X_{ij}) = \phi e^{\alpha_i + \beta_j + c}$. Our aim is to estimate $\alpha$ and $\beta$ parameters using maximum likelihood method under the usual constraint that $\alpha_1 = \beta_1 = 0$.

\subsection{Gamma Model}

On the one hand, the model described in \cite[3.3]{EV2002} is the following. Increments are assumed to be Gamma distributed random variables, i.e., $X_{ij} \sim \Gamma (\alpha, \beta)$, with expected value $E(X_{ij}) = m_{ij}$ and variance $Var(X_{ij}) = \phi m_{ij}^2$. Thus parameters are $\alpha = \frac{1}{\phi}$ and $\beta = \frac{1}{\phi m_{ij}}$.

On the other hand, in \cite[5.2.4]{MV2008} $X_{ij}$ increments are deterministic sums of ($r_{ij}$) independent Gamma random variables. In other words, $X_{ij} = \sum \limits_{k=1}^{r_{ij}} X^{(k)}_{ij}$, where $X^{(k)}_{ij} \sim \Gamma (\nu, \frac{\nu}{m_{ij}})$. Since the rate parameters are equal, the distribution of $X_{ij}$ is $\Gamma(\nu r_{ij}, \frac{\nu}{m_{ij}})$. Here, $E(X_{ij}) = r_{ij} m_{ij}$ and $Var(X_{ij}) = \frac{r_{ij}}{\nu} m^2_{ij}$.

If $r_{ij} = 1 ~~ \forall i,j$, the above mentioned two models are identical, and $\nu = \frac{1}{\phi}$. The only difference is that $r_{ij}$'s can be arbitrary integers, thus the second model is more flexible. Note that for handling estimation methods, it is not satisfying to know the upper triangle, but the triangle of $r_{ij}$ numbers too. Model Assumptions 5.19 in \cite{MV2008} state that there exist $\mu_1, \ldots, \mu_n$ and $\gamma_1, \ldots, \gamma_n$ parameters under the constraint that $\sum \limits_{i=1}^n \gamma_i = 1$, such that $E(X_{ij}) = r_{ij} m_{ij} = \mu_i \gamma_j$. The estimates $\hat{\mu}_i$ and $\hat{\gamma}_j$ are averages of the observations weighted by numbers $r_{ij}$. For more details see Model Assumptions 5.19. in the book.

On the one hand, $\hat{\mu}$ and $\hat{\gamma}$ parameters are the solution of the following system of equations:

$$\hat{\mu}_i = \frac{\sum \limits_{j=1}^{n+1-i} \frac{X_{ij}}{\hat{\gamma}_j}}{n+1-i} ~~ , ~~ \hat{\gamma}_j = \frac{\sum \limits_{i=1}^{n+1-j} \frac{X_{ij}}{\hat{\mu}_i}}{n+1-j}.$$

(Note that for technical reasons, in our simulations we used simple chain-ladder method instead of solving the above mentioned system of equations.) On the other hand, the estimation of parameter $\nu$ using Pearson residuals is the following. For $i+j \leq n+1$ let
$$\hat{r}^P_{ij} = \frac{x_{ij} - \hat{\mu}_i \hat{\gamma}_j}{\hat{\mu}_i \hat{\gamma}_j},$$ thus
$$\hat{\phi}_P = \frac{\sum \limits_{i+j \leq n+1} \left( \hat{r}^P_{ij} \right)^2}{\frac{n(n+1)}{2} - (2n-1)}.$$

\section[Stochastic Claims Reserving]{Stochastic Claims Reserving}
\label{sec:reserve}
\subsection{Parametric models}
We are using the parametric models introduced previously. We estimate the parameters from the upper triangle, and approximate the conditional distribution of the lower triangle by determining the conditional distribution with the estimated parameters. Naturally, this approach might have drawbacks, for example the prediction intervals will not be precise - to see the problem, we could just think of how to make a prediction interval for the $(n+1)$th member of an i.i.d. sample of normal distribution from the first $n$ members.
As the conditional distribution of the lower triangle is difficult, and in some cases, impossible to calculate analitically, we estimate the distribution with Monte Carlo method, generating 5000 lower triangles.

\subsection{Bootstrap methods with over-dispersed Poisson and gamma distributions}
The detailed description of these models can be found in \cite{EV2002}, Appendix 3. These models are also included in the \texttt{ChainLadder} package for \texttt{R}. The heart of these methods is bootstrapping the adjusted Pearson residuals $r_{i,j}^{adj}= \sqrt{\frac{n}{1/2\cdot n(n+1)-2n+1}}\cdot r_{i,j}^{P}$ of the upper triangle. From the resampled residuals, we create a new upper triangle, fit the standard chain-ladder to it, obtain the new expected values $\widetilde{ m}_{i,j}$ and variances $\phi \widetilde{ m}_{i,j}$ for the elements of the lower triangle, simulate the lower triangle, and store the ultimate claim amount. We do the bootstrapping 5000 times, and use the 5000 stored ultimate claim amounts as the predictive distribution.

\subsection{Semi-stochastic methods}
We use the models presented in \cite{FVA2007}. Consider $\alpha _{j}\left ( i \right )= \frac{C_{i,j+1}}{C_{i,j}}$ , and let $\alpha _{j}, j=1,...,n$ be independent discrete uniform random variables with $P\left ( \alpha _{j}= \alpha _{j}\left ( i \right ) \right )= \frac{1}{n-j}$. The model assumption is that for each element of the lower triangle, the $C_{i,j+1}=C_{i,j}\cdot  \alpha _{j}$ equation holds. It is easy to see that $E\left ( \alpha _{j} \right )= \frac{1}{n-j}\sum \limits_{i=1}^{n-j}\frac{C_{i,j+1}}{C_{i,j}}$.
The first method (denoted as Uniform) is to simulate a sufficiently large number (in our case, 5000) of lower triangles, take the arousing 5000 ultimate claims, and use this empirical distribution as the predicted distribution.
The second method (denoted as Unifnorm) is based on the assumption that the ultimate claim amount is nearly a normal random variable, and the task is to calculate the mean and variance. The mean is the expected value of the ultimate claim amount, which is $\sum \limits_{j=1}^{n}\left ( C_{n-j+1,j}\prod \limits_{k=j}^{n-1}E\left ( \alpha _{k} \right ) \right )$. \\The variance of the ultimate claim is $\sum \limits_{j=1}^{n}\left ( C_{n-j+1,j}\left ( \prod \limits_{k=j}^{n-1}E\left ( \alpha_{k}^{2} \right )-\prod \limits_{k=j}^{n-1}E^{2}\alpha _{k} \right ) \right )$.

\section[Probabilistic Forecasts]{Probabilistic Forecasts}
\label{sec:score}

\subsection{Introductory Example}

As we mentioned in the previous section, stochastic reserving is essentially the prediction of the distribution of the future payments. The systematic analysis of these distributions begun with the work of \cite{D1984}. Most applications of probabilistic forecast theory is used in meteorology. \cite{GBR2007} give an overwiev of the most important methods for comparing the forecasts. To present these methods, we give 2 simple examples, the first of which is similar to the one in the aforementioned article.

In \emph{Example 1}, suppose that an automobile insurance company already has paid 1 million euros for the accidents occured in year 2012. Let $\xi$ be the total amount paid for accidents of 2012 till the end of 2013. Assume that $\xi$ is of log-normal distribution with log-scale and shape parameters $\mu$ and 1, respectively, where $\mu$ is supposed to be a standard normal random variable.

In the integer valued \emph{Example 2}, the number of liability insurance claims for damages occured and reported in 2012 was 1000. Let $\eta$ be the number of claims for damages regarding 2012, that are reported in 2013. Assume that the distribution of $\eta$ is Poisson with parameter $1000 \cdot \lambda$, where lambda is a gamma distributed random variable with shape parameter $1.5$ and scale parameter $0.5$.

We compare the performance of 4 imaginary actuaries predicting the distributions in the above examples.
The ideal actuary knows all the relevant circumstances, which means the knowledge of the exact value of $\mu$ in \emph{Example 1}, and $\lambda$ in \emph{Example 2}, respectively.
The long-term actuary does not intend to get involved in the actual information, and simply estimates the unconditional distribution.
The ordinary actuary attempts to estimate the parameters, but the estimation has some possible error.
At last, the intern actuary finds the expected value, but does not care about the distribution itself.
We give an overview of the distributions and the predicted distributions in Table \ref{tab:ex1table} and Table \ref{tab:ex2table}. Year 2013 was simulated $10,000$ times and the different probabilistic forecasts were compared according to them.

\begin{table}
\begin{center}
\begin{tabular}{ll} \hline \hline
actuary & predictive distribution \\ \hline \hline
 \rowcolor[gray]{.92} ideal & $LN(\mu, 1)$ \\
long-term & $LN(0,2)$ \\
 \rowcolor[gray]{.92} ordinary & $\frac12 LN(\mu, 1) + \frac12 LN(\mu + \delta, 1)$, where $\delta = \pm 1$ with probability $\frac12, \frac12$ \\
intern & $LN(-|\mu|, \sigma^2)$, where $-|\mu|+\sigma^2 = \mu + \frac12$, i.e. $\sigma^2 = \begin{cases} 4\mu + 1 & \mu \geq 0 \\ 1 & \mu < 0 \end{cases}$ \\ \hline \hline
\end{tabular}
\end{center}
\caption{Distributions regarding certain actuaries in Example 1 (Log-normal). $\xi \sim LN(\mu, 1)$, where $\mu \sim N(0,1)$.} \label{tab:ex1table}
\end{table}

\begin{table}
\begin{center}
\begin{tabular}{ll} \hline \hline
actuary & predictive distribution \\ \hline \hline
 \rowcolor[gray]{.92} ideal & Poisson$(x \cdot \lambda)$ \\
long-term & NegBin$(1.5, \frac{1}{1+x \cdot 0.5})$ \\
 \rowcolor[gray]{.92} ordinary & $\frac12 $Poisson$(x \cdot \lambda) + \frac12 $Poisson$(x \cdot \lambda \cdot \delta)$, where $\delta = 1 \pm \frac{1}{10}$ with probability $\frac12, \frac12$ \\
intern & NegBin$(2x \cdot \lambda, \frac23)$ \\ \hline \hline
\end{tabular}
\end{center}
\caption{Distributions regarding certain actuaries in Example 2 (Poisson). $\eta \sim$Poisson$(x \cdot \lambda)$, where $\lambda \sim \Gamma(1.5, 0.5)$ and $x=1000$.} \label{tab:ex2table}
\end{table}

\subsection{PIT, Empirical Coverage, Average Width}

In the evaluation of the predictions, we have $(F_i,x_i)$ pairs, where $F_i$ denotes the predicted distribution in the $i$th case, and $x_i$ the realization, respectively. \cite{D1984} suggested using the \emph{Probability Integral Transform} (PIT), and the description available in \cite{GBR2007} is also worth seeing. The main idea is that substituting a random variable with continuous probability function into its own distribution function, will yield a uniformly distributed variable on the (0,1) interval. The distribution of PIT values has to be uniform on interval $(0,1)$, otherwise the estimation is biased. Remark that the uniform rank histogram property is a \emph{necessary} but not sufficient criterion for ideal forecasts, as illustrated by an example in \cite{H2000}. For a fixed upper triangle - as a condition - we calculate the predicted distribution function of the ultimate claim, and substitute the real ultimate claim into it.

We check the prediction by plotting the empirical CDF of the corresponding PIT values and comparing it to the identity function, see P-P plots below, or by plotting the histogram of the PIT values and checking for uniformity. On the one hand, U-shaped histograms indicate underdispersed, excessively light-tailed predictive distributions. On the other hand, $\cap$-shaped histograms hint at overdispersion, too heavy predictive distribution in tails, and skewed histograms occur when central tendencies are biased.

\begin{figure}[tbp]
\centering
\includegraphics[width=.6\textwidth]{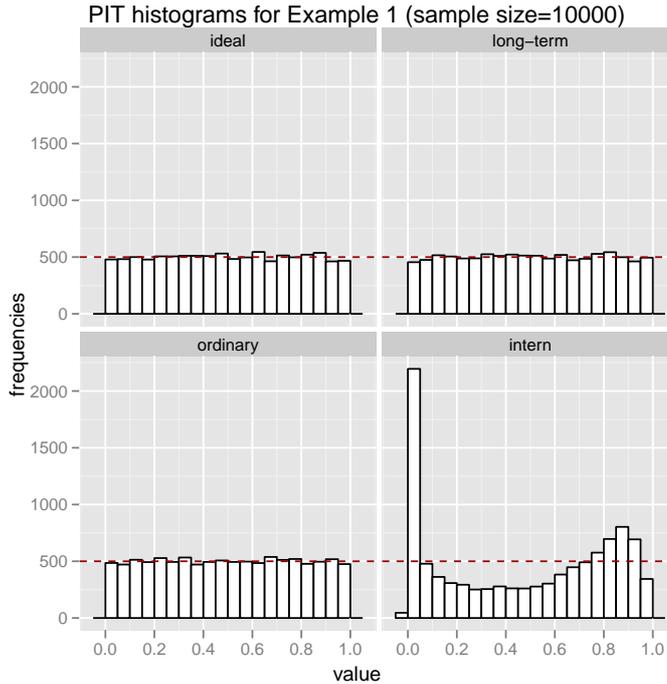}
\caption{PIT histograms for Example 1 (Log-normal example)}
\label{fig:ex1pit}
\end{figure}

\begin{figure}[tbp]
\centering
\includegraphics[width=.6\textwidth]{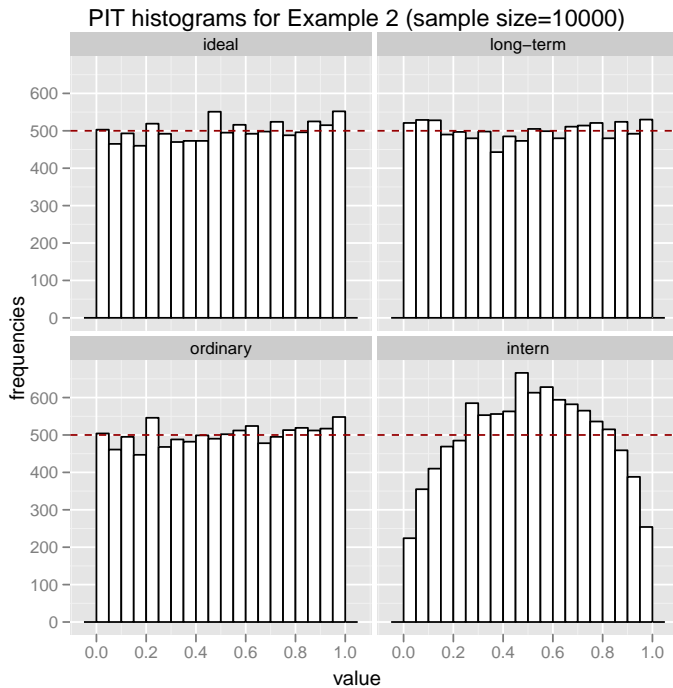}
\caption{PIT histograms for Example 2 (Poisson example)}
\label{fig:ex2pit}
\end{figure}

In high count data cases this method can be used very well. In low count data cases, randomized PIT, or non-randomized uniform version of the PIT histogram is the more appropriate choice, see \cite{CGH2009}. PIT histograms regarding the two example can be seen on Figure \ref{fig:ex1pit} and Figure \ref{fig:ex2pit}, where dashed lines represent uniform distributions. These figures provide additional examples for inaccurate probabilistic forecasts with appropriate PIT histograms, since shapes regarding long-term and ordinary predictions deviate not more considerable from uniform distribution, as in the ideal case. Nevertheless, the intern case shows inappropriateness immediately.

We also intend to define P-P plots for random variables $\xi$ and predictive distributions $\hat{F}$. Namely, let $s_p := \sup \limits_{z} \{ z : \hat{F}(z)\leq p \}$, thus the P-P plot function ($[0,1] \to [0,1]$) is as follows: $p \mapsto P(\xi < s_p)$. Remark that integrating the PIT function results the P-P plot, i.e., there is a bijection between the two concepts. It also follows obviously that a slanted-S shaped P-P plot corresponds to a $\cap$-shaped PIT, for instance. An example for run-off triangles can be seen on Figure \ref{fig:raaquan}. Using the notations of our simulations, for a real number $p \in (0,1)$, the function value is $\frac1N \sum \limits_{j=1}^N \chi_{\{ \xi_j < Qu(\eta_{\cdot j}, p) \}}$, where $Qu(\eta_{\cdot j}, p)$ is the $p$-quantile value of empirical distribution defined by $\eta_{1j}, \ldots, \eta_{Mj}$.

At last, two linked measures will be considered for each probabilistic forecast. The idea derives from \cite{BNH2012}, and an example calculation can be seen on Table \ref{tab:raacover}. Coverage in \% related to a central prediction interval $\alpha \cdot 100\%$ is the proportion of observations between the lower and upper $\frac{1-\alpha}{2}$ quantiles. In accordance with our terminology, it can be written as
$\frac1N \sum \limits_{j=1}^N \chi_{\{ Qu(\eta_{\cdot j}, (1-\alpha)/2) < \xi_j < Qu(\eta_{\cdot j}, (1+\alpha)/2 ) \}}$. The average width for a central prediction interval is the difference of lower and upper $\frac{1-\alpha}{2}$ quantiles expressed in payments, in expectation. It can be interpreted as the sharpness of the prediction. If the performances of two stochastic claims reserving techniques are relatively close to each other, measured in scores, the one with the narrower width is the better. Formally we calculated the following way:
$\frac{1}{N} \sum \limits_{j=1}^N \left( \eta^*_{\frac{(1+\alpha)\cdot M}{2}, j} - \eta^*_{\frac{(1-\alpha)\cdot M}{2}, j} \right)$. Even though these could be derived from the PIT, in some applications it is important to know the exact values, thus a separate calculation is advised. See Table \ref{tab:ex1cover} and Table \ref{tab:ex2cover} regarding \emph{Example 1} and \emph{Example 2}, respectively.
It can be seen again, looking at the mentioned tables that inappropriate distributions may provide good results. Namely, the coverage of the long-term and ordinary actuary, and the average width of the intern actuary is acceptable as well. However, in our examples only the ideal gives suitable values for both measures.

\begin{table}[h] \small
\begin{center}
\begin{tabular}{l rr rr}
\hline \hline
\multicolumn{1}{c}{\multirow{3}{*}{Actuary}}
        & \multicolumn{2}{c}{\multirow{2}{*}{Coverage (\%)}}
        & \multicolumn{2}{c}{\multirow{2}{*}{Average width}}\\
\multicolumn{1}{c}{\multirow{2}{*}{}}
        & \multicolumn{4}{c}{\multirow{2}{*}{ }}\\ \cline{2-5}
 & 66\% & 90\% & 66\% & 90\% \\ 
  \hline
\hline
ideal & 67.0 & 90.5 & 1.93 & 3.29 \\ 
   \rowcolor[gray]{.92} long-term & 67.3 & 90.5 & 2.74 & 4.65 \\ 
  ordinary & 67.2 & 90.4 & 5.01 & 11.64 \\ 
   \rowcolor[gray]{.92} intern & 47.3 & 74.2 & 2.87 & 4.88 \\ 
  
\hline \hline
\end{tabular}
\caption{Coverage and Average width in Example 1 (Log-normal)}
\label{tab:ex1cover}
\end{center}
\end{table}

\begin{table}[h] \small
\begin{center}
\begin{tabular}{l rr rr}
\hline \hline
\multicolumn{1}{c}{\multirow{3}{*}{Actuary}}
        & \multicolumn{2}{c}{\multirow{2}{*}{Coverage (\%)}}
        & \multicolumn{2}{c}{\multirow{2}{*}{Average width}}\\
\multicolumn{1}{c}{\multirow{2}{*}{}}
        & \multicolumn{4}{c}{\multirow{2}{*}{ }}\\ \cline{2-5}
 & 66\% & 90\% & 66\% & 90\% \\ 
  \hline
\hline
ideal & 66.2 & 89.5 & 48.91 & 83.16 \\ 
   \rowcolor[gray]{.92} long-term & 65.5 & 89.5 & 1052.00 & 1868.00 \\ 
  ordinary & 66.4 & 89.5 & 98.24 & 140.62 \\ 
   \rowcolor[gray]{.92} intern & 76.2 & 95.2 & 59.89 & 101.84 \\ 
  
\hline \hline
\end{tabular}
\caption{Coverage and Average width in Example 2 (Poisson)}
\label{tab:ex2cover}
\end{center}
\end{table}

\subsection{Continuous Ranked Probability Score and Energy Score}

Probability scores are an increasingly widespread technique to measure the quality of the predicted distributions. A score is a function of the predictive distribution and a realization of the real value. The predictive distribution can be represented by its CDF, empirical CDF or PDF, depending on which score is used. Forecast models are then ranked by comparing the average score of the predictive distributions from each model.

The most highlighted measure of goodness of fit in this article is the so called \emph{Continuous Ranked Probability Score} or CRPS, which is devoted to handle the case of prediction of distribution functions.
The main reason of giving preference to the CRPS over other several score concepts is the general applicability, i.e., it can be applied to predictions (functions) regardless of absolute continuity or discrete behaviour.
A detailed description can be found in \cite[p. 366.]{GT2007}, for instance. Briefly, let $F$ denote the predictive distribution function, and $c$ one single observation. Here

$$CRPS(F,c) = - \int \limits_{-\infty}^{\infty} (F(y) - 1_{\{y \geq c\}})^2 dy.$$

(See Table \ref{tab:raasummarytab} and Figure \ref{fig:raacrps}.)
A generalization of CRPS is the so called \emph{Energy Score}, see the aforementioned \cite{GT2007}, for instance. The definition of one dimensional energy score is
$$ES(F,c) = \frac12 E_F |X-X'|^\beta - E_F |X-c|^\beta,$$
where $X,X'$ are i.i.d. variables from distribution $F$, and $\beta \in (0,2)$. It can be shown that for $\beta = 1$ we get the CRPS back.
We note that CRPS can be evaluated directly, or in case of $\beta \neq 1$, it is viable to approximate the energy score by sampling from the empirical distribution, which is typically a much less rapid way, because of the necessary sample size to be drawn to reach a satisfying accuracy. (During our simulations, the magnitude of difference in time was approximately $10^2$, although even so not significant.)

Table \ref{tab:ex12score} presents the score results in expectation, regarding the two examples (Log-normal and Poisson). For instance, consider the Log-normal example in case of the ideal actuary. Given $\{ \mu = m \}$ and $\{ \xi = x \}$, the score value is in the form $\int_0^\infty \left( \Phi(\log y - m) - 1_{\{ y \geq x \}} \right)^2 dy$.
It turns out that the score measure provided proper ranking, although, the difference between ideal and intern values regarding the second example is not of distinction.

\begin{table}[ht]
\begin{center}
{\small
\begin{tabular}{lrrrr}
  \hline
\hline
 & ideal & long-term & ordinary & intern \\
  \hline
\hline
Example 1 & -0.48 & -1.86 & -0.63 & -1.83 \\
  Example 2 & -14.49 & -327.62 & -32.62 & -14.64 \\
   \hline
\hline
\end{tabular}
}
\caption{CRPS results in Example 1 and Example 2}
\label{tab:ex12score}
\end{center}
\end{table}

\subsection{Mean square error of prediction (MSEP)}

The mean square error of prediction (MSEP) of predictor (point estimation) $\hat{X}$ for ultimate claim $X$, conditionally on the $\sigma$-algebra $\mathcal{F}$ is defined as
$$msep_{X|\mathcal{F}} (\hat{X}) = E\left( \left( X-\hat{X} \right)^2 |\mathcal{F} \right) = Var(X|\mathcal{F}) + (\hat{X} - E(X|\mathcal{F}))^2.$$
See Definition 3.1 in \cite{MV2008}, for instance. The unconditional MSEP is
$$ msep_{X} (\hat{X}) =E\left( \left( X-\hat{X} \right)^2 \right)= E(Var(X|\mathcal{F})) + E((\hat{X} - E(X|\mathcal{F}))^2).$$ In the detailed examples the above defined errors can be calculated right away.

For instance, actuaries according to Example 2 have mean square error of predictions as follows.

$$msep_{\eta}(\text{ideal}) = msep_{\eta}(\text{intern}) = E(msep_{{\eta|\lambda }} (x\lambda)) =$$ $$= E(Var(\eta|\lambda)) + E((x\lambda - E(\eta|\lambda))^2)=E(x\lambda)=x\cdot \alpha\cdot \beta$$

$$msep_{\eta}(\text{long-term}) = msep_{\eta}(E(\eta)) =Var(\eta)=\frac{\alpha \left( 1-\frac1{1+x\cdot \beta} \right)}{ \left( \frac1{1+x\cdot \beta} \right)^2}= x\cdot \alpha\cdot \beta \cdot (1+x\cdot \beta )$$

$$msep_{\eta}(\text{ordinary}) = E(msep_{\eta|\lambda } (\frac1{2}\cdot x\lambda\cdot (1+\delta))) =$$ $$= E(Var(\eta|\lambda)) + E((\frac1{2}\cdot x\lambda\cdot (1+\delta)- E(\eta|\lambda))^2)= $$
$$=x\cdot \alpha\cdot \beta +\frac{x^2}{400}\cdot E(\lambda^2)=x\cdot \alpha\cdot \beta \left(1+\frac{x\cdot(1+\alpha)\cdot \beta}{400} \right)$$

where $\alpha=1.5$, $\beta=0.5$ denote the shape and scale parameters of the gamma distributions. Results in conjunction with the two simple example are shown on Table \ref{tab:ex12msep}. The tables show that in the long-term case, the prediction is inaccurate even though the PIT histogram and the prediction interval coverages suggest that the fit is good.

\begin{table}[ht]
\begin{center}
{\small
\begin{tabular}{lrrrr}
  \hline
\hline
 & ideal & long-term & ordinary & intern \\
  \hline
\hline
Example 1 & 34.5 & 47.2 & 37.2 & 34.5 \\
  Example 2 & 750.0 & 375750.0 & 3093.8 & 750.0 \\
   \hline
\hline
\end{tabular}
}
\caption{Mean Square Error of Prediction in Example 1 and Example 2}
\label{tab:ex12msep}
\end{center}
\end{table}
When we work with run off triangles, the definiton is the following. $$msep_{UC|\mathcal{D}_j} (\hat{UC}_j) = Var(UC|\mathcal{D}_j) + (\hat{UC}_j - E(UC|\mathcal{D}_j))^2 ~~~~ j=1,\ldots, N,$$ for each real triangle.
Let $x_j$ be the \emph{real} ultimate claims ($j=1,\ldots,N$) and $y_{ij}$ the \emph{generated} ultimate claims ($i=1,\ldots,M$, $j=1,\ldots,N$). On the other hand, let $z_{ij}$ be an ultimate claim generated with the real parameters and according to the real development distribution, conditionally on upper triangle $\mathcal{D}_j$. Thus, in our calculations
$$msep_{UC|\mathcal{D}_j} (\hat{UC}_j) \approx \frac{\sum \limits_{i=1}^M (z_{ij} - E(UC|\mathcal{D}_j))^2}{M-1} + \left( \frac{\sum \limits_{i=1}^M y_{ij}}{M} - \frac{\sum \limits_{i=1}^M z_{ij}}{M} \right)^2.$$
Remark that if we calculate the msep values supposing the knowledge of real parameters instead of estimation, we get the pure $Var(UC|\mathcal{D}_j)$ back, since $y_{ij} = z_{ij}$.

On the other hand, Table \ref{tab:raasummarytab} shows MSEP values regarding the run-off triangle example detailed in Section \ref{sec:simul}.

\section[Simulation and Results]{Simulation and Results}
\label{sec:simul}




\subsection{Preliminary Remarks}

Our calculations have been implemented in $\texttt{R}$, and beyond a self made program code, we used the $\texttt{ChainLadder}$ package. The documentation of the latter can be found on webpage \\ \url{http://cran.r-project.org/web/packages/ChainLadder/ChainLadder.pdf}. The former is available on url \url{http:// on request} with a short user manual enclosed. Remark that the detailed simulation results for several parameter sets are also to find on this page. Moreover, the completion of tables and figures have been made using packages $\texttt{xtable}$ and $\texttt{ggplot2}$.

In this section, a Monte Carlo type method will be introduced, followed by simulated examples, consisting of corresponding goodness of fit values described in Section \ref{sec:score}. Parameters of the example come from the run-off triangle RAA, which is an accumulated claims triangle from the Automatic Facultative business in General Liability, originally published in Historical Loss Development, Reinsurance Association of America (RAA), 1991, and was also used as an example in \cite{EV2002} and \cite{M1994}. It is crucial to emphasize that this well known triangle plays no role in the present article apart from providing parameters. In other words, during our research work we fitted several distributional models only to get parameter values from real data, but without studying goodness of fits, which is another objective. Due to lack of space, only a special example is presented, but additional simulation results are available on the aforementioned webpage. Just to mention some claims data from literature, these are related to ABC (a workers' compensation portfolio of a large company, first used in \cite{BZ2000}, which has an in-depth analysis of the data array as well), GenIns (a general claims data triangle from \cite{TA1983}, and was also used in \cite{M1993}) and M3IR5 (a simulated triangle, from \cite{Z1994}).

\subsection{Monte Carlo Type Method}

This subsection is devoted to the detailed description of our Monte Carlo type method (MC method) for comparison of various claims reserving methods in case of different distributional background of run-off triangles. In other words, our goal is to establish a ranking among the different stochastic claims reserving techniques, if the real development property of claims payments for accident years is in accordance with certain models, as described in Section \nameref{sec:model}, for instance. This comparison now is not based on mean square error of prediction or quantile values, but on scoring rules. Nevertheless, the results and plots containing the former values are also included in the article, which allows us to compare them with the unconventional score results in theory of insurance. Remark that the important role in our approach is played by ultimate claim values, i.e., the aggregate $\sum \limits_{i=1}^n C_{i,n}$ payments for accident years $1, \ldots, n$. Although, if someone is more interested in focusing on solely the next year payments, i.e., in $\sum \limits_{j=2}^n X_{n-j+2,j}$, the Monte Carlo type method can be interpreted easily without any special alteration. This latter reserving value is suggested by Solvency II.

Suppose that the development distribution model of run-off triangle and corresponding parameters are given. As a \emph{first step}, we generate $N$ run-off triangles independently, and besides, for each the ultimate claim values $UC_1, UC_2, \ldots, UC_N$. Let $\mathcal{D}_1, \mathcal{D}_2, \ldots, \mathcal{D}_N$ denote these upper triangles, as conditions to be used later.

In the \emph{second step}, for every generated upper triangle, we calculate the predicted distribution of the ultimate claims, using the methods described in \nameref{sec:model}. As we mentioned there, we determine these predicted distributions via Monte Carlo method. Specifically, in the parametric models case, the parameters for each generated upper triangle have to be evaluated. They will certainly differ from each other, and from the real parameters as well. Assuming these parameter values for each condition $\mathcal{D}_i$, or upper triangle, in other words, $M$ ultimate claim values have to be generated, denoted by $\hat{UC}_{i,1}, \hat{UC}_{i,2}, \ldots, \hat{UC}_{i,M}$, as stochastic predictors. Thus we get the predictive distributions. Following that, we prepare the forecasts using the 2 bootstrap, the uniform and the uniform normal methods.

At last, in the \emph{third step}, for each pair $UC_i$ and tuple $(\hat{UC}_{i,1}, \hat{UC}_{i,2}, \ldots, \hat{UC}_{i,M})$ scores, PITs, msep and quantile values, and additionally the confidence intervals (coverage and average width) have to be calculated according to Section \nameref{sec:score}. Generally, a predictive distribution function $F_i$ derives from values $\hat{UC}_{i,1}, \hat{UC}_{i,2}, \ldots, \hat{UC}_{i,M}$, and the corresponding $c$ value is $UC_i$.

As an example, the results for regarding a Gamma Model example are shown on several figures below. $N$ is chosen to be 2000, and $M$ to be 5000. Figure \ref{fig:raapit} contains the histograms of 2000 PIT values. On the one hand, Figure \ref{fig:raacrps} consists of boxplot representations of CRPS scores for each claims reserving technique. On the other hand, the mean CRPS values can be found on Table \ref{tab:raasummarytab}. Similarly, Figure \ref{fig:raaensc} represents the energy scores for $\beta = \frac12$. Remember that the higher score values mean more appropriate reserving methods. Remark that on boxplots, the 'reserving methods' axis has to be interpreted as follows:
\begin{center}
\begin{tabular}{llll}
1 - Log-normal & 2 - Negative Binomial & 3 - Poisson \\
4 - Over-dispersed P. & 5 - Gamma & 6 - Uniform \\
7 - Unif. Normal & 8 - Bootstrap Gamma & 9 - Bootstrap Od. Pois. \\
10 - Ideal &&
\end{tabular}
\end{center}
We have not mentioned method 'Ideal' so far, which differs from all other methods essentially. Namely, for each distributional model, we assume that parameters are known, and use them in the above described second step, instead of any estimation. The reason of the inclusion of these results is that theoretically this means the best way of the prediction of claims reserves, in expectation. In other words, we used it for reference purposes, recall the example of ideal actuary.

\subsection{Results on a Gamma distributed example}

We now present the results for a Gamma distributional model. These can be interpreted in a number of different ways.
\begin{enumerate}
\item Which score or error number is the most consistent, and how do they correlate with each other? Do the equipments applied in stochastic predictions choose the actual models better than regular measures, such as mean square error of prediction, for instance?
\item Which non-parametric, distributional free methods predict the distributions properly? Do they outperform prediction methods derived from parametric models?
\item How reliable and sharp are the prediction intervals?
\end{enumerate}
Parameters are
$$(\mu_1, \ldots, \mu_{10}) = (21048, 17507, 23723, 29562, 25751, 18680, 15676, 22141, 19019, 18402),$$
$$(\gamma_1, \ldots, \gamma_{10}) = (0.112, 0.224, 0.209, 0.147, 0.119, 0.092, 0.037, 0.031, 0.016, 0.009),$$ and $\nu = 2.22$.
Based on the MSEP values on Table \ref{tab:raasummarytab}, the first thing that stand out is the poor fit of the Log-normal model. The other 4 parameter estimating models are roughly the same. The bootstrap methods are slightly worse, and the uniform and unifnorm methods are giving very poor results compared to the others.

\begin{figure}[tbp]
\centering
\includegraphics[width=.6\textwidth]{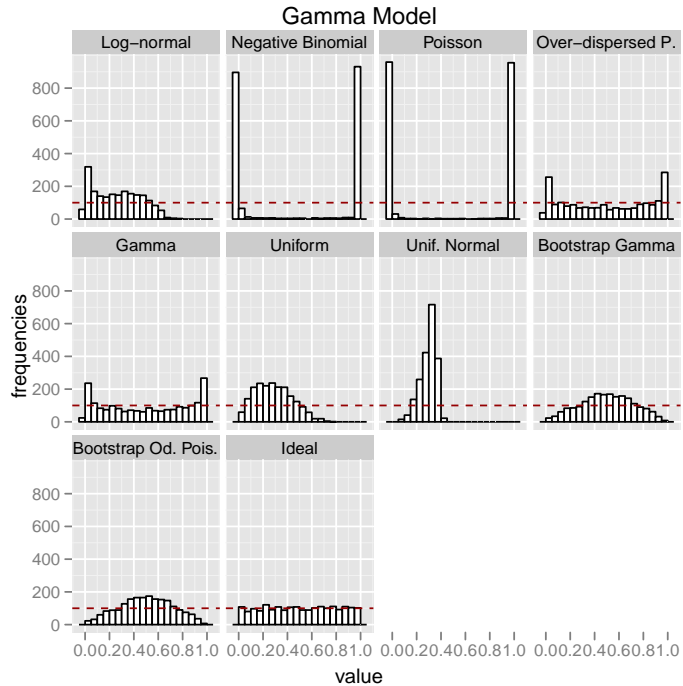}
\caption{Histograms of PIT values in the Gamma distributed example}
\label{fig:raapit}
\end{figure}

\begin{table}[ht]
\begin{center}
{\small
\begin{tabular}{lrrrr}
  \hline
\hline
Res. Method & CRPS (mean) & En. Sc. (mean) & MSEP (mean) & MSEP (median) \\
  \hline
\hline
Log-normal & -17840 &  -77.47 & 6.257e+09 & 2.159e+09 \\
   \rowcolor[gray]{.92} Negative Binomial &  -9878 &  -81.03 & 1.799e+08 & 9.856e+07 \\
  Poisson & -10010 &  -84.23 & 1.799e+08 & 9.862e+07 \\
   \rowcolor[gray]{.92} Over-dispersed P. &  -7547 &  -57.10 & 1.800e+08 & 9.911e+07 \\
  Gamma &  -6911 &  -54.28 & 1.539e+08 & 9.109e+07 \\
   \rowcolor[gray]{.92} Uniform & -20980 &  -80.25 & 6.144e+09 & 4.716e+09 \\
  Unif. Normal & -50040 & -145.90 & 6.139e+09 & 4.711e+09 \\
   \rowcolor[gray]{.92} Bootstrap Gamma &  -7856 &  -57.82 & 2.021e+08 & 9.880e+07 \\
  Bootstrap Od. Pois. &  -7865 &  -57.84 & 2.015e+08 & 1.005e+08 \\
   \rowcolor[gray]{.92} Ideal &  -4074 &  -41.53 & 5.296e+07 & 5.295e+07 \\
   \hline
\hline
\end{tabular}
}
\caption{Scores and Mean Square Errors in the Gamma Model example}
\label{tab:raasummarytab}
\end{center}
\end{table}
If we take a look at the PIT histograms on Figure \ref{fig:raapit}, we can see that the Negative Binomial and Poisson distributional model produces very poor fits, as we might have expected. In other words, these provide great examples for light-tailed predictions. The reason is that the model parameters imply e.g. a Poisson variable with a very high expected value, which makes the standard deviation very low compared to it. Hence the ultimate claim values will be in a very narrow range compared to the predicted distributions, leading to PIT histograms, where all the occurrences are in a 10-20 percent wide probability range of the predicted distributions. The other side of this attribute of the Poisson distribution is that when we use a Poisson distribution as the predicted distribution, the difference between the smallest and the largest value of the empirical predicted distribution will be very small, consequently the real ultimate claim values will most of the time be either bigger or smaller than all 5000 values of the predicted distribution. Thus, the corresponding PIT graphs will contain occurrences almost exclusively in the 5 percent and 95 percent probability levels. In essence, the PIT analysis strongly suggests against using the Poisson distribution as the incremental claims of the IBNR claims in the current example, as even though the MSEP values are acceptable, the low variance attribute leads to the Poisson distribution seemengly being little upgrade over a simple point estimation of the expected value. Although, it is important to remark that considering real data, occurrence of nearly Poisson distributed triangles for pay amounts is not unprecedented. Looking at the $\cap$-shaped histograms deriving from the bootstrap methods, there is an example for heavy-tailed predictors to be seen.

\begin{figure}[h]
\centering
\includegraphics[width=.6\textwidth]{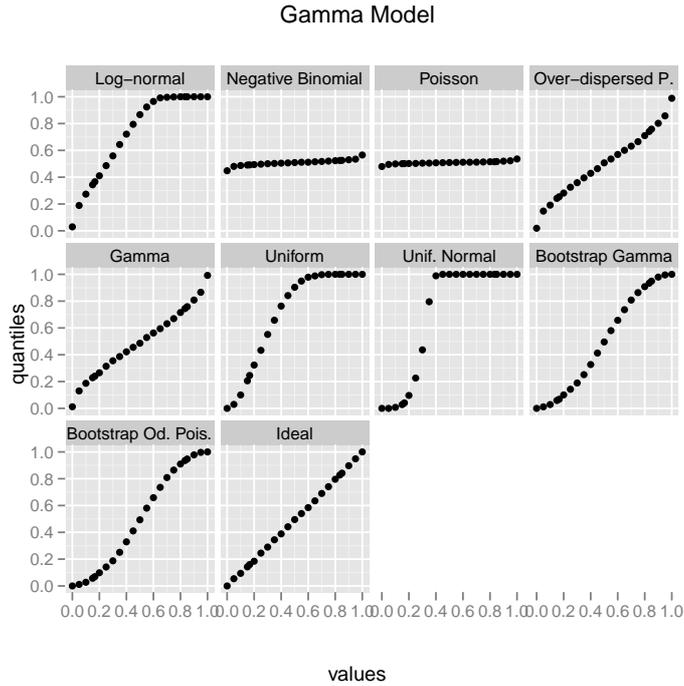}
\caption{P-P Plots in the Gamma Model example}
\label{fig:raaquan}
\end{figure}

When examining the P-P plots, we expected to see the best results, i.e., a graph closest to the identity function, when the real distribution is of the same type as the one in the MC method, and the calculations proved just that, see Figure \ref{fig:raaquan}. Although, even if we get the Gamma model right, but do not know the real parameters, the prediction is slightly overdispersed.
We can conclude that in terms of the consistency of quality of the P-P plots, the bootstrap methods provide quite acceptable results. The Uniform and Uniform normal semi-stochastic methods, even though we expected them to perform almost as well as the bootstrap methods in the P-P plot test, are giving worse results. Regarding the Poisson distribution, we can come to the same conclusions as we did in the PIT analysis: it provides too little variance to effectively predict any other distribution, and predicting it with any other method, the variance of the resulting distribution would be too big to be considered a good fit.

\begin{figure}[h]
\centering
\includegraphics[width=.6\textwidth]{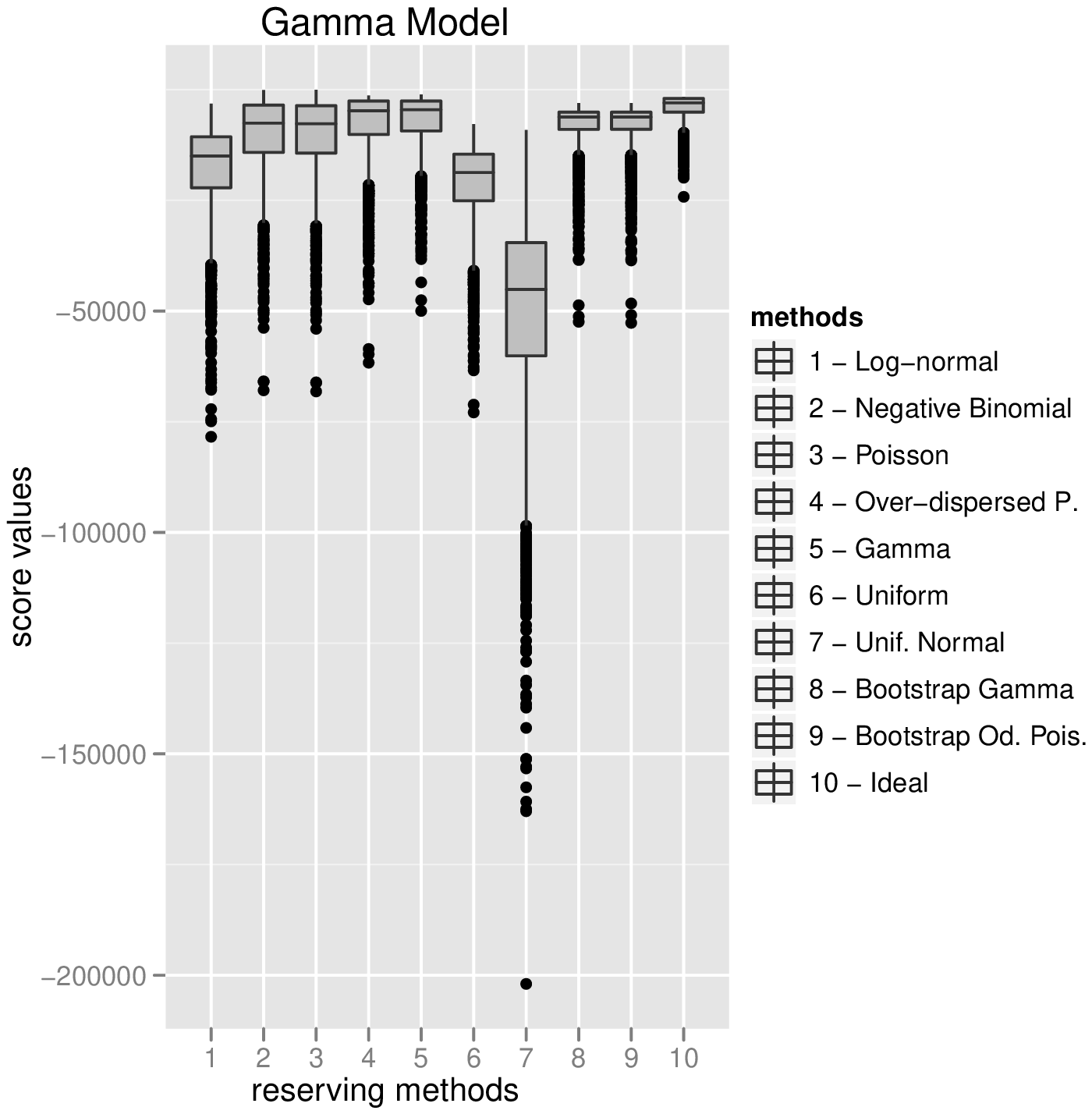}
\caption{Boxplots of CRPS values in the Gamma Model example}
\label{fig:raacrps}
\end{figure}

\begin{figure}[h]
\centering
\includegraphics[width=.6\textwidth]{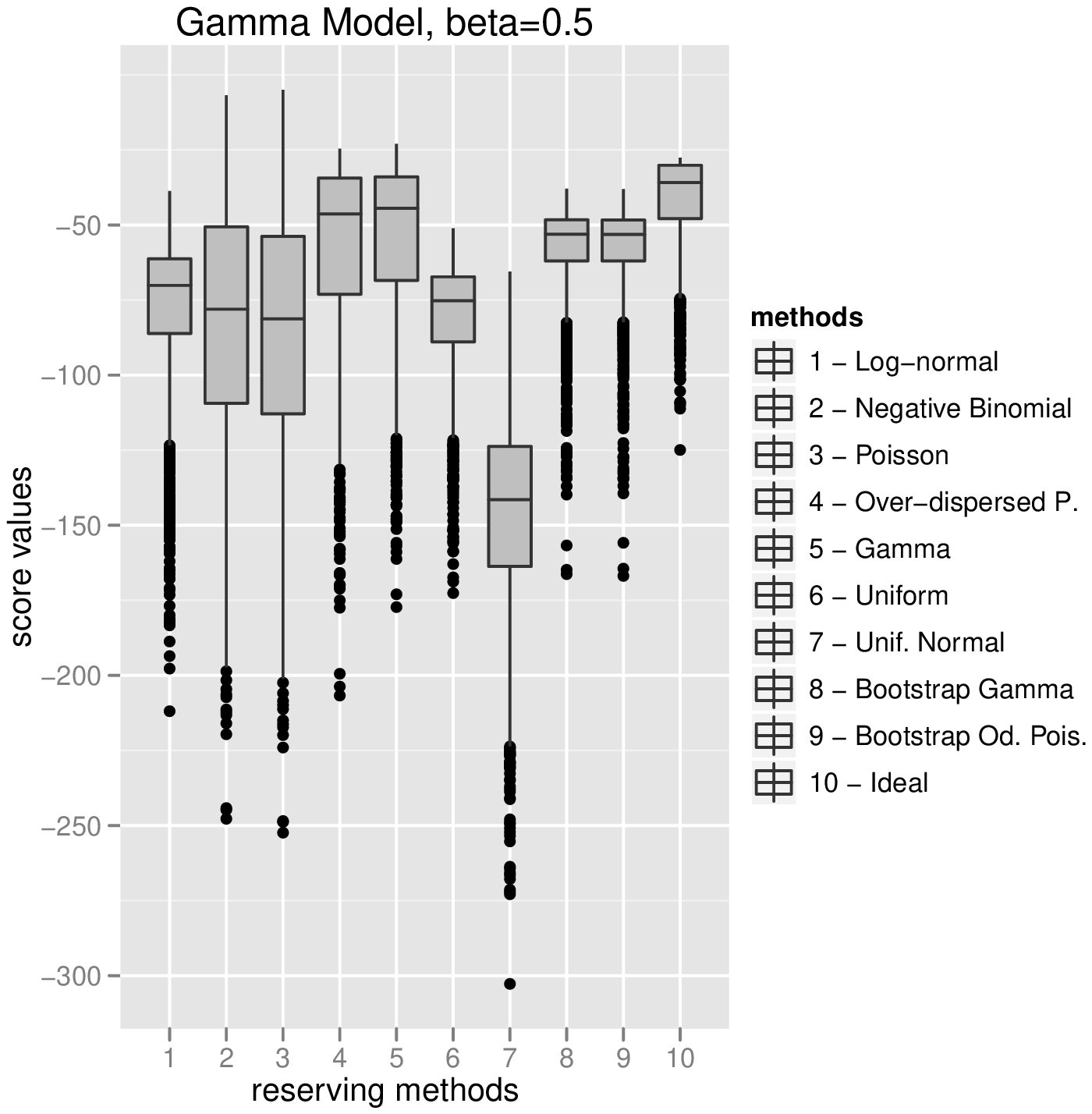}
\caption{Boxplots of Energy Score values in the Gamma Model example}
\label{fig:raaensc}
\end{figure}

Regarding the stochastic methods in the example, the Gamma model is clearly the best in the score metric, followed by Negative Binomial, and bootstrap methods.
See Figure \ref{fig:raacrps} and \ref{fig:raaensc}, which contain the boxplot representations of CRPS and Energy Score values. Table \ref{tab:raasummarytab} summarizes the mean score values.

On the other hand, the Uniform and Uniform normal methods give even worse results than if we were using a wrong distribution in any of the 5 MC methods, so even though those 2 do not require us to guess the underlying distribution, their results are worse then if we were making the wrong guess, and then build a parameter approximating MC model based on that guess.

Table \ref{tab:raacover} presents the inaccuracy of the applied prediction intervals, moreover, the other run-off triangles and models gave similar results in this sense. In cases where the coverage is said to be acceptable, the sharpness is mostly inappropriate, i.e., intervals are wide. The coverage values for Log-normal model are closer to the $66\%$ and $90\%$ values, then in the Gamma case, but average width results are higher.

\begin{table}[h] \small
\begin{center}
\begin{tabular}{l rr rr}
\hline \hline
\multicolumn{1}{c}{\multirow{3}{*}{Reserving method}}
        & \multicolumn{2}{c}{\multirow{2}{*}{Coverage (\%)}}
        & \multicolumn{2}{c}{\multirow{2}{*}{Average width}}\\
\multicolumn{1}{c}{\multirow{2}{*}{}}
        & \multicolumn{4}{c}{\multirow{2}{*}{ }}\\ \cline{2-5}
 & 66.67\% interval & 90\% interval & 66.67\% interval & 90\% interval \\ 
  \hline
\hline
Log-normal & 63.4 & 81.1 & 89035 & 245659 \\ 
   \rowcolor[gray]{.92} Negative Binomial & 3.3 & 5.4 & 919 & 1562 \\ 
  Poisson & 1.4 & 2.8 & 444 & 755 \\ 
   \rowcolor[gray]{.92} Over-dispersed P. & 48.7 & 71.0 & 15497 & 26322 \\ 
  Gamma & 50.4 & 73.6 & 15506 & 26533 \\ 
   \rowcolor[gray]{.92} Uniform & 75.4 & 97.0 & 111512 & 422161 \\ 
  Unif. Normal & 95.9 & 100.0 & 287986 & 489479 \\ 
   \rowcolor[gray]{.92} Bootstrap Gamma & 86.6 & 98.4 & 39946 & 70131 \\ 
  Bootstrap Od. Pois. & 86.6 & 98.5 & 39951 & 70112 \\ 
   \rowcolor[gray]{.92} Ideal & 66.8 & 89.4 & 13967 & 23821 \\ 
  
\hline \hline
\end{tabular}
\caption{Coverage and Average width in the Gamma Model example}
\label{tab:raacover}
\end{center}
\end{table}



\subsection{Public payments data example}

In case of the detailed dataset provided by an insurance company, we proceeded not exactly as in the analysis of the known run-off triangles. Which means, each of the 2000 real run-off triangles were generated the following way. We used sampling with replacement $43,081$ times from the set of $43,081$ accidents, which determines a triangle. Moreover, it resulted in a $6 \times 6$ quadrangle also, and in the real ultimate claim, respectively. Following this simulation, we did exactly as described before, i.e., determined the predictive distributions for the 2000 triangles with several methods, and compared them to the real ultimate claim values.

\begin{figure}[tbp]
\centering
\includegraphics[width=.6\textwidth]{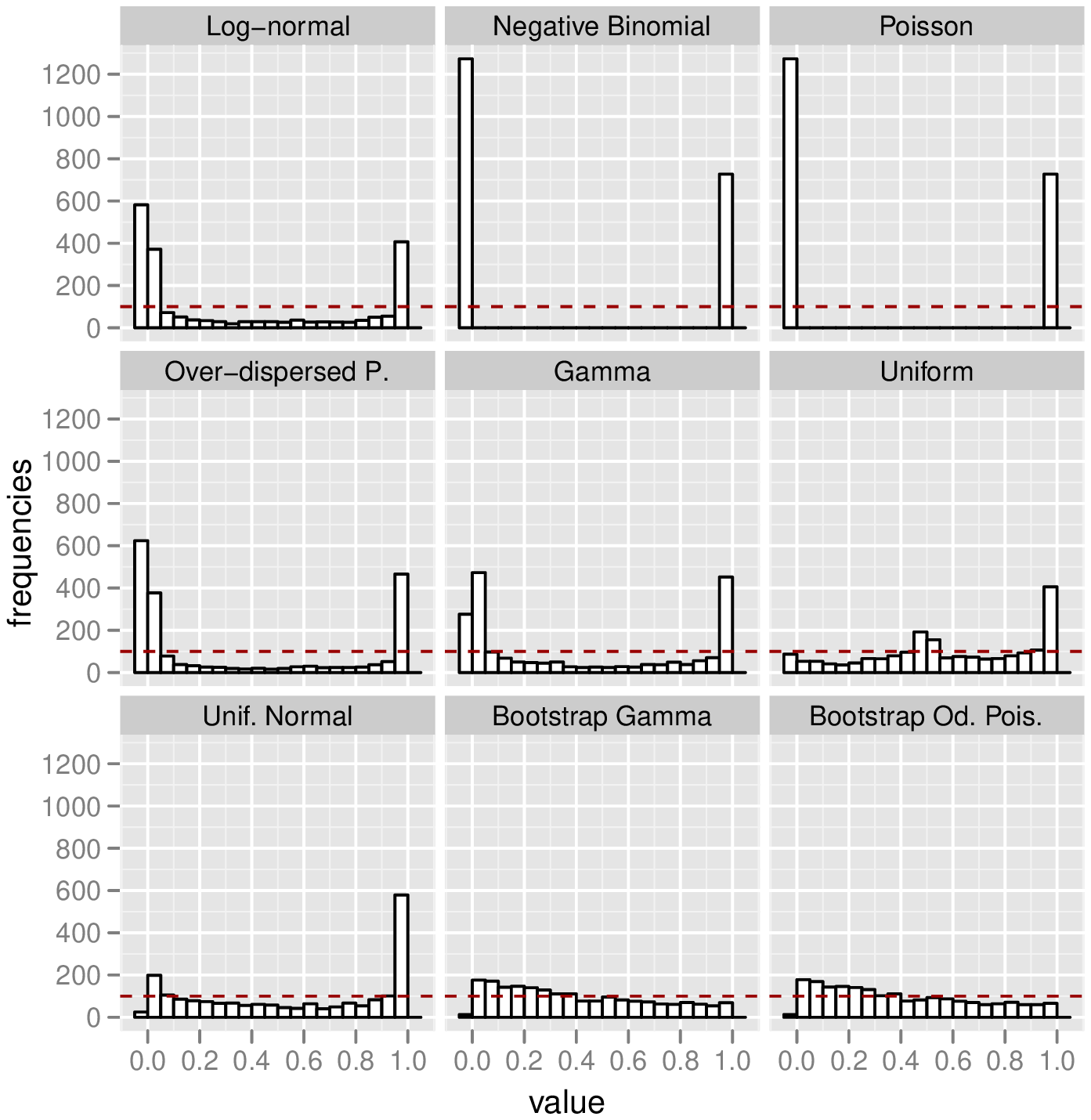}
\caption{Histograms of PIT values in the Public Data case}
\label{fig:pubpit}
\end{figure}

Figure \ref{fig:pubpit} shows that in the light of PIT values, none of the methods based on distributional models are recommended. Best histograms are resulted by the bootstrap estimation methods. Although uniform and uniform normal methods are not much worse, they underperform in the cases of extreme claim values. It is also confirmed by Table \ref{tab:pubcover}, i.e., the proportion of prediction intervals of bootstrap methods are close to the $66.67\%$ and $90\%$ values. Unfortunately, in exchange for the appropriate coverage percentages, average width values are higher.

\begin{table}[h] \small
\begin{center}
\begin{tabular}{l rr rr}
\hline \hline
\multicolumn{1}{c}{\multirow{3}{*}{Reserving method}}
        & \multicolumn{2}{c}{\multirow{2}{*}{Coverage (\%)}}
        & \multicolumn{2}{c}{\multirow{2}{*}{Average width}}\\
\multicolumn{1}{c}{\multirow{2}{*}{}}
        & \multicolumn{4}{c}{\multirow{2}{*}{ }}\\ \cline{2-5}
 & 66.67\% interval & 90\% interval & 66.67\% interval & 90\% interval \\ 
  \hline
\hline
Log-normal & 19.2 & 31.9 & 1261338486 & 2170190302 \\ 
   \rowcolor[gray]{.92} Negative Binomial & 0.0 & 0.0 & 573564 & 974878 \\ 
  Poisson & 0.0 & 0.0 & 163897 & 278579 \\ 
   \rowcolor[gray]{.92} Over-dispersed P. & 15.2 & 26.7 & 954272655 & 1621537701 \\ 
  Gamma & 23.8 & 40.0 & 1395724439 & 2375100112 \\ 
   \rowcolor[gray]{.92} Uniform & 56.2 & 72.5 & 3228640816 & 4738399784 \\ 
  Unif. Normal & 38.8 & 59.9 & 1963981053 & 3337926040 \\ 
   \rowcolor[gray]{.92} Bootstrap Gamma & 61.5 & 87.1 & 4490491606 & 8679547627 \\ 
  Bootstrap Od. Pois. & 61.6 & 87.1 & 4492743160 & 8679552105 \\ 
  
\hline \hline
\end{tabular}
\caption{Coverage and Average width in the Public Data case}
\label{tab:pubcover}
\end{center}
\end{table}

On the one hand, according to the CRPS values in Table \ref{tab:pubsummarytab}, the uniform claims reserving method has the highest score, which has not been performing well for other triangles. On the other hand, energy scores are slightly better for the bootstrap methods. It has to be mentioned that the msep values are the highest in the bootstrap cases, in other words, this measure implies a different ranking.

\begin{table}[ht]
\begin{center}
{\small
\begin{tabular}{lrrrr}
  \hline
\hline
Reserving Method & CRPS (mean) & En. Sc. (mean) & MSEP (mean) & MSEP (median) \\
  \hline
\hline
Log-normal & -1.526e+09 & -27780 & 5.420483e+18 & 3.699599e+18 \\
   \rowcolor[gray]{.92} Negative Binomial & -1.817e+09 & -38850 & 5.348926e+18 & 3.671239e+18 \\
  Poisson & -1.817e+09 & -38970 & 5.348924e+18 & 3.671239e+18 \\
   \rowcolor[gray]{.92} Over-dispersed P. & -1.579e+09 & -28820 & 5.350275e+18 & 3.671087e+18 \\
  Gamma & -1.397e+09 & -25880 & 4.803373e+18 & 3.239738e+18 \\
   \rowcolor[gray]{.92} Uniform & -1.227e+09 & -23530 & 4.101087e+18 & 2.726758e+18 \\
  Unif. Normal & -1.231e+09 & -23600 & 4.100103e+18 & 2.723995e+18 \\
   \rowcolor[gray]{.92} Bootstrap Gamma & -1.347e+09 & -23470 & 7.204460e+19 & 5.300120e+18 \\
  Bootstrap Od. Pois. & -1.347e+09 & -23470 & 4.082055e+19 & 5.308000e+18 \\
   \rowcolor[gray]{.92}  \hline
\hline
\end{tabular}
}
\caption{Scores and Mean Square Errors in the Public Data case}
\label{tab:pubsummarytab}
\end{center}
\end{table}

\section[Conclusions]{Conclusions}
\label{sec:conc}

We conclude our article by trying to answer the 3 questions proposed in the previous section. The method that gives the best results overall, and is the least sensitive to the underlying distribution is the bootstrap gamma and bootstrap over-dispersed Poisson method. Had we chosen a different one for our paper, we might have deduced that the 2 bootstrap methods are the same in strength. It also became clear that the Uniform and Uniform normal methods are significantly worse than the bootstraps in every evaluation. The 5 parameter approximating MC methods give good results when we guess the underlying distribution correctly, and as the P-P plots and PIT values show, they can be much worse then the bootstraps when applied to the wrong distribution.
The PIT and P-P plots advised against the use of the Poisson model for the RAA dataset. When we studied other datasets, we found that the Poisson model exhibits the same limitations, however, the Log-normal model was just as good as the other distributions. In general, scores provide a reasonable fit to the background distribution. Unfortunately, the methods used in this paper -- and also applied in insurance industry -- result not reliable prediction intervals.

This negative result is not surprising, since a typical run-off triangle contains much less information than the unknown parameters. To improve the predictions, \cite{Me2007} proposed the usage of Bayesian methods. In our opinion, other than these methods, it would also be worthwhile to try using individual contract and claim data in probabilistic forecasts.

In our work, we used the individual claim dataset of an insurance company as well, but not for probabilistic forecasting, but to propose a technique for comparing stochastic methods. This technique is not sufficiently mathematically established yet, but may be applied in practice. One can simulate many scenarios from the past, and choose a model, which best fits the claims data of the insurance institution.

We hope our work helped to show that in stochastic claims reserving, both in theoretical and in applied situations, it is worthwhile to test the quality of the different methods, and in multiple ways if possible.

As it was mentioned before, our goal was not the construction of the best stochastic claims reserving technique, but to propose an adequate methodology for comparisons in the future. We hope to take the first step in this direction with our work.

\cleardoublepage \phantomsection
\renewcommand{\bibname}{References}
\bibliographystyle{plainnat}
\bibliography{biblio}

\begin{thebibliography}{22}
\providecommand{\natexlab}[1]{#1}
\providecommand{\url}[1]{\texttt{#1}}
\expandafter\ifx\csname urlstyle\endcsname\relax
  \providecommand{\doi}[1]{doi: #1}\else
  \providecommand{\doi}{doi: \begingroup \urlstyle{rm}\Url}\fi

\bibitem[Baran et~al.(2012)Baran, Nemoda, and Hor\'{a}nyi]{BNH2012}
S.~Baran, D.~Nemoda, and A.~Hor\'{a}nyi.
\newblock Probabilistic wind speed forecasting in {H}ungary.
\newblock arXiv:1202.4442, 2012.

\bibitem[Barnett and Zehnwirth(2000)]{BZ2000}
G.~L. Barnett and B.~Zehnwirth.
\newblock Best estimates for reserves.
\newblock In \emph{Proceedings of the Casualty Actuarial Society LXXXVII},
  2000.

\bibitem[Bj\"{o}rkwall et~al.(2009)Bj\"{o}rkwall, H\"{o}ssjer, and
  Ohlsson]{BHO2009}
S.~Bj\"{o}rkwall, O.~H\"{o}ssjer, and E.~Ohlsson.
\newblock Non-parametric and parametric bootstrap techniques for age-to-age
  development factor methods in stochastic claims reserving.
\newblock \emph{Scandinavian Actuarial Journal}, 2009\penalty0 (4):\penalty0
  306--331, 2009.
\newblock \doi{10.1080/03461230903239738}.
\newblock URL
  \url{http://www.tandfonline.com/doi/abs/10.1080/03461230903239738}.

\bibitem[Czado et~al.(2009)Czado, Gneiting, and Held]{CGH2009}
C.~Czado, T.~Gneiting, and L.~Held.
\newblock Predictive model assessment for count data.
\newblock \emph{Biometrics}, 65\penalty0 (4):\penalty0 1254--1261, 2009.
\newblock ISSN 1541-0420.
\newblock \doi{10.1111/j.1541-0420.2009.01191.x}.
\newblock URL \url{http://dx.doi.org/10.1111/j.1541-0420.2009.01191.x}.

\bibitem[Dawid(1984)]{D1984}
A.~P. Dawid.
\newblock Present position and potential developments: Some personal views:
  Statistical theory: The prequential approach.
\newblock \emph{Journal of the Royal Statistical Society. Series A (General)},
  147\penalty0 (2):\penalty0 pp. 278--292, 1984.
\newblock ISSN 00359238.
\newblock URL \url{http://www.jstor.org/stable/2981683}.

\bibitem[England and Verrall(2002)]{EV2002}
P.~D. England and R.~J. Verrall.
\newblock Stochastic claims reserving in general insurance.
\newblock \emph{British Actuarial Journal}, 8\penalty0 (3):\penalty0 443--518,
  2002.

\bibitem[England and Verrall(2006)]{EV2006}
P.~D. England and R.~J. Verrall.
\newblock Predictive distributions of outstanding liabilities in general
  insurance.
\newblock \emph{Annals of Actuarial Science}, 1\penalty0 (02):\penalty0
  221--270, 8 2006.
\newblock ISSN 1748-5002.
\newblock \doi{10.1017/S1748499500000142}.
\newblock URL \url{http://journals.cambridge.org/article_S1748499500000142}.

\bibitem[Faluk\"{o}zy et~al.(2007)Faluk\"{o}zy, Vit\'{e}z, and
  Arat\'{o}]{FVA2007}
T.~Faluk\"{o}zy, I.~I. Vit\'{e}z, and M.~Arat\'{o}.
\newblock \emph{Stochastic models for claims reserving in insurance business},
  chapter~13, pages 102--113.
\newblock World Scientific, 2007.
\newblock \doi{10.1142/9789812709691_0013}.
\newblock URL
  \url{http://www.worldscientific.com/doi/abs/10.1142/9789812709691_0013}.

\bibitem[Gneiting and Raftery(2007)]{GT2007}
T.~Gneiting and A.~E. Raftery.
\newblock Strictly proper scoring rules, prediction, and estimation.
\newblock \emph{Journal of the American Statistical Association}, 102\penalty0
  (477):\penalty0 359--378, Mar 2007.
\newblock URL \url{http://dx.doi.org/10.1198/016214506000001437}.

\bibitem[Gneiting et~al.(2007)Gneiting, Balabdaoui, and Raftery]{GBR2007}
T.~Gneiting, F.~Balabdaoui, and A.~E. Raftery.
\newblock Probabilistic forecasts, calibration and sharpness.
\newblock \emph{Journal of the Royal Statistical Society: Series B (Statistical
  Methodology)}, 69\penalty0 (2):\penalty0 243--268, 2007.
\newblock ISSN 1467-9868.
\newblock \doi{10.1111/j.1467-9868.2007.00587.x}.
\newblock URL \url{http://dx.doi.org/10.1111/j.1467-9868.2007.00587.x}.

\bibitem[Hamill(2001)]{H2000}
T.~M. Hamill.
\newblock Interpretation of rank histograms for verifying ensemble forecasts.
\newblock \emph{Monthly Weather Review}, 129\penalty0 (3):\penalty0 550--560,
  March 2001.
\newblock URL
  \url{http://nldr.library.ucar.edu/repository/collections/AMS-PUBS-000-000-000-008}.

\bibitem[Mack(1993)]{M1993}
T.~Mack.
\newblock Distribution-free calculation of the standard error of chain ladder
  reserve estimates.
\newblock \emph{ASTIN Bulletin -- The Journal of the International Actuarial
  Association}, 23\penalty0 (2):\penalty0 213--225, November 1993.
\newblock URL \url{http://www.actuaries.org/LIBRARY/ASTIN/vol23no2/213.pdf}.

\bibitem[Mack(1994)]{M1994}
T.~Mack.
\newblock Which stochastic model is underlying the chain ladder method?
\newblock \emph{Insurance: Mathematics and Economics}, 15\penalty0
  (2-3):\penalty0 133--138, 1994.
\newblock URL
  \url{http://EconPapers.repec.org/RePEc:eee:insuma:v:15:y:1994:i:2-3:p:133-138}.

\bibitem[M\'{a}rkus and Gyarmati-Szab\'{o}(2007)]{GM2007}
L.~M\'{a}rkus and J.~Gyarmati-Szab\'{o}.
\newblock \emph{A Hierarchical Bayesian model to predict belatedly reported
  claims in insurances}, chapter~17, pages 137--144.
\newblock World Scientific, 2007.
\newblock \doi{10.1142/9789812709691_0017}.
\newblock URL
  \url{http://www.worldscientific.com/doi/abs/10.1142/9789812709691_0017}.

\bibitem[Merz and W\"{u}thrich(2010)]{MV2010}
M.~Merz and M.~V. W\"{u}thrich.
\newblock Paid-incurred chain claims reserving method.
\newblock \emph{Insurance: Mathematics and Economics}, 46\penalty0
  (3):\penalty0 568--579, 2010.
\newblock URL
  \url{http://EconPapers.repec.org/RePEc:eee:insuma:v:46:y:2010:i:3:p:568-579}.

\bibitem[Meyers(2007)]{Me2007}
G.~Meyers.
\newblock Thinking outside the triangle, paper presented to the 37th astin
  colloquium, florida, 2007.

\bibitem[Pinheiro et~al.(2003)Pinheiro, e~Silva, and Centeno]{PSC2003}
P.~J.~R. Pinheiro, J.~M.~A. e~Silva, and Maria de~Lourdes Centeno.
\newblock Bootstrap methodology in claim reserving.
\newblock \emph{The Journal of Risk and Insurance}, 70\penalty0 (4):\penalty0
  701--714, 2003.
\newblock ISSN 00224367.
\newblock URL \url{http://www.jstor.org/stable/3519936}.

\bibitem[Taylor and Ashe(1983)]{TA1983}
G.~C. Taylor and F.~R. Ashe.
\newblock Second moments of estimates of outstanding claims.
\newblock \emph{Journal of Econometrics}, 23\penalty0 (1):\penalty0 37--61,
  September 1983.
\newblock URL \url{http://ideas.repec.org/a/eee/econom/v23y1983i1p37-61.html}.

\bibitem[Taylor(2000)]{T2000}
G.C. Taylor.
\newblock \emph{Loss Reserving: An Actuarial Perspective}.
\newblock Huebner International Series on Risk, Insurance and Economic Security
  Series. Springer-Verlag GmbH, 2000.
\newblock ISBN 9780792385028.
\newblock URL \url{http://books.google.hu/books?id=NmM-EH4h7kcC}.

\bibitem[Verrall(1994)]{V1994}
R.~J. Verrall.
\newblock A method for modelling varying run-off evolutions in claims
  reserving.
\newblock \emph{ASTIN Bulletin -- The Journal of the International Actuarial
  Association}, 24\penalty0 (2):\penalty0 325 -- 332, November 1994.
\newblock URL \url{http://www.casact.org/library/astin/vol24no2/325.pdf}.

\bibitem[W\"{u}thrich and Merz(2008)]{MV2008}
M.~V. W\"{u}thrich and M.~Merz.
\newblock \emph{Stochastic Claims Reserving Methods in Insurance}.
\newblock Wiley, 2008.

\bibitem[Zehnwirth(1994)]{Z1994}
B.~Zehnwirth.
\newblock Probabilistic development factor models with applications to loss
  reserve variability, prediction intervals, and risk based capital.
\newblock In \emph{Variability in Reserves Prize Program Papers}, volume~2.
  Casualty Actuarial Science Forum, 1994.

\end{thebibliography}

\end{document}